\documentstyle[12pt]{article}
\begin{document}
\def\singlespacing{\baselineskip=12pt}
\def\doublespacing{\baselineskip=18pt}
\doublespacing

\pagestyle{empty}

\begin{center}
\begin{Large}
{\bf Theoretical and numerical studies of chemisorption
on a line with precursor layer diffusion} \\
\end{Large}
\bigskip
\bigskip
J.\ A.\ N.\ Filipe\footnote{email: Joao.Filipe@brunel.ac.uk}
and G.\ J.\ Rodgers\footnote{email: Rodgers@ph.brunel.ac.uk} \\
\bigskip
Department of Physics, Brunel University, Uxbridge, Middlesex UB8 3PH, UK \\
\bigskip
\bigskip
June 13, 95 \\
Submitted to Phys.\ Rev.\ E

\bigskip
\bigskip
\bigskip
\bigskip
\bigskip

{\bf Abstract}
\end{center}
\bigskip
We consider a model for random deposition of monomers on a line with
extrinsic precursor states.
As the adsorbate coverage increases, the system develops
non-trivial correlations due to the diffusion mediated deposition
mechanism.
In a numeric simulation, we study various quantities describing
the evolution of the island structure.
We propose a simple, self-consistent theory which incorporates
pair correlations.
The results for the correlations, island density number, average
island size and probabilities of island nucleation, growth and
coagulation show good agreement with the simulation data.

\bigskip
\bigskip
\bigskip

\noindent arch-ive/9508039

\newpage

\pagestyle{plain}
\pagenumbering{arabic}

\section{Introduction}
\label{s:intro}

Models of {\it random sequential adsorption} (RSA) have been used to describe
the process by which particles are irreversibly deposited without overlap
onto a surface.
They are relevant to studying the adsorption of gas molecules \cite{book} or
colloid particles in solution \cite{gas.diffusion} onto solid surfaces,
or of large molecules on biological membranes \cite{membranes}.
One of the key assumptions, is that the particles bind strongly to the
substrate so that desorption and surface mobility are negligible on the
time-scale of the experiment.
Although providing a very simplified picture, RSA models have the virtue
of being exactly solvable in dimension $d=1$, for monomer and for
$k$-mer particles \cite{1dsolution}.

Many additional features have been added to this model in order to make it
more physical, or to generalize its range of applications, and the literature
in the field is vast \cite{review}.
The inclusion of diffusional relaxation on the substrate
\cite{surf.diff,surf.diff-desorp}, leading to equilibration, or of cooperative
effects, such as multi-site exclusion \cite{nn.exclusion}, have been examined.
The possibilities that the particles are reflected back to the fluid or
desorbed \cite{surf.diff-desorp,HTW,King}, or that the surface comprises
more than one chemical species \cite{XK} have also been considered.

Another important possibility, motivated by many experimental studies of
gas-metal surfaces \cite{experiment,KW}, is that if the particles lose just
enough of their kinetic energy they can become trapped (physisorbed) in a
mobile, temporary {\it precursor state}, from which they can be adsorbed
(chemisorbed) at a far site at a later time.
The mechanism of precursor mediated chemisorption, first postulated by Taylor
and Langmuir \cite{precursor}, was initially formulated as a statistical model
by Kisliuk \cite{Kisliuk}. It was later adapted by King and coauthors
\cite{King,KW,CK,book} to include other effects, such as temperature
dependence,
desorption, molecular dissociation and pair interactions between the adatoms.
Variations on the Kisliuk model have also been studied by other authors, both
numerically and analytically \cite{HTW,XK,WBH,AD,BB,Evans,EN}.
Most analytical treatments, however, are essentially mean-field like, in that
they largely ignore the correlations between the state of occupation of
different sites which arise through the precursor mediated deposition process,
i.e.\ they assume that the rate of deposition at a given site only depends on
the state of occupation of that site, and not on the neighbouring sites.

In this paper, we consider a lattice model which is both a simplified version
of the Kisliuk model, and a slight complication of the RSA model.
Contrary to most previous studies, we examine the case where the mobility of
physisorbed atoms is only possible on the top of occupied chemisorption sites
({\it extrinsic} precursor states), and where there is neither reflection nor
desorption back to the gas phase (later we shall comment on how scattering and
desorption can in principle be included into the equations of motion).
Each deposition attempt will, therefore, result either in the direct occupation
of an empty site, or in the diffusion on the top of an island of occupied sites
until the particle finds an empty site at the edge of the island, where it is
irreversibly deposited.
For simplicity, we also consider that the deposited particles are
{\it monomers}, or atoms, in which case the lattice will eventually become
full.
As a consequence of the (extrinsic) precursor diffusion, the edges of islands
are preferential sites for chemisorption and the growth of the larger islands
is favoured.
This introduces non-trivial correlations between sites, especially at
late-times
when diffusion is the dominant deposition mechanism, and makes the model
unsolvable even for the simplest case of monomer deposition on a {\it line}.
A model similar to the one adopted here has been examined by Becker
and Ben-Shaul in $2d$ \cite{BB}.
Their analysis, however, only applies to the early kinetics, as they treat the
islands as uncorrelated objects, completely neglecting the contribution of
island coalescence to the growth process.

Regarding the motivation for our study, we recall a somewhat unrealistic
feature of the RSA model, namely that once a deposition attempt fails, the
particle's position is 'randomized' before another adsorption attempt is made.
As the transport of particles to the surface is diffusive, one might expect
that a failed deposition attempt would be followed by another nearby adsorption
attempt \cite{no.randomize}.
Since the intrinsic precursor seems to play no role in this effect, the present
model may account for it, at least partially.
Furthermore, it has been previously suggested by Cassuto and King \cite{CK},
that a kinetic deposition model without intrinsic precursor states and with
negligible desorption, could well also explain the experimental data from some
gas-solid systems, such as hydrogen on tungsten.
Naively, one may also think that such a model could be able describe the slow
deposition of liquid droplets on a plane surface, the spread of epidemics from
immunized to non-immunized populations, or the growth of a forest where seeds
have to be transported to an open field to find suitable conditions to develop.

Although island formation and structure kinetics have been studied
(see e.g.\ \cite{nn.exclusion,EN}), most work on RSA and precursor models,
either analytical or simulation, has focused on determining the dependence of
the coverage fraction $\theta$ and sticking probability on the exposure,
i.e.\ on the time $t$, and its value at the {\it jamming limit} (which is
trivial only for monomers).
In the present model, we assume that the diffusion time scale is small enough
compared to the time scale of deposition attempts, and so the coverage
is proportional to time or, equivalently, the sticking probability is 1.
We examine the case of a substrate of dimension $d=1$, and concentrate on
studying the quantities describing the evolving morphology of the occupied
regions. Namely, the pair correlation functions, the total number of
islands, and the probability distribution of the island sizes and its moments.
We have measured these quantities in numerical simulations, and developed a
minimal theory incorporating correlations, to calculate them approximately.
The results show good agreement, while at the same time differing considerably
from those of the RSA model for monomer deposition in $1d$ where there are no
spatial correlations at all.

A brief summary of this paper is the following.
In section 2 we describe the model and define the formalism.
The results of the numerical simulation are discussed in section 3.
The theory is presented in section 4, and the results are compared
with the simulation data. Section 5 is dedicated to some conclusions.

\section{Model}
\label{s:model}

We consider a one dimensional lattice with $N$ sites, with $N$ large enough
to neglect boundary effects, i.e.\ we take $N\to\infty$ in the calculations.
At each site $i$ we define a variable $S_i$, such that
\begin{eqnarray}
S_i & = & 0  \ \ \ {\rm if \ the \ site \ is \ empty}   \nonumber \\
    & = & 1  \ \ \ {\rm if \ the \ site \ is \ occupied} \nonumber \ .
\end{eqnarray}
Every time step $\delta t=1$ a site is chosen randomly and a monomer
deposition attempt is made.
If the site is empty the particle is adsorbed and the site becomes
occupied irreversibly.
Otherwise, the particle diffuses on top of the occupied region until it
reaches an empty site where it is adsorbed (this is appropriately modelled
by a random walk with traps \cite{random.walk}). Only then is the next
deposition attempted. The process repeats until the lattice is full.

We assume here that the particle diffusion is rapid enough to be over before
the time of the next deposition attempt, independently of the size of
the island where it takes place.
This may sound as an unrealistic assumption, especially if the island is of the
order of the system size. We note, however, that due to the randomness of the
process it is unlikely that a diffusing particle would interact or compete with
the next particle to be deposited. This argument fails, of course, in the limit
when the lattice is almost full (intermediate coagulation may occur) or in the
case when diffusion is very slow (a gas of net precursor particles develops).

Since each time step results in a deposition, either direct or mediated by
diffusion, the time dependence of the adsorbate coverage,
$\theta=\left<S_i\right>$, is trivial in this case:
\begin{equation}
\theta(t) = t/N \ \ , \ \ 0\leq t \leq N \ . \label{cov.D=1}
\end{equation}

One can easily incorporate in the model the possibility of {\it scattering}
of a particle back into the gas phase at the instant of collision with
the surface. If the scattering probability is $(1-\alpha)$, whatever
the state of occupation of the site, the rate of adsorption per unit time
must be multiplied by $\alpha$. We may then set $\alpha =1$ through a
redefinition of the time scale. To account for the possibility that scattering
may depend on the state of occupation of the site, we set the probability of
no chemisorption to be $(1-D\alpha)$ if a particle lands on an occupied site.
In this case, the rate of variation of the number of occupied sites is
(setting $\alpha=1$)
\begin{equation}
Nd\theta/dt = 1(1-\theta) + D\theta = 1 - (1-D)\theta \ , \label{cov.eq}
\end{equation}
with initial condition $\theta(0)=0$. The solution is
\begin{equation}
\theta(t) = \frac{1-\exp\left[-(1-D)t/N\right]}{1-D} \ . \label{cov.D}
\end{equation}
This reduces to (\ref{cov.D=1}) for $D=1$, and to the RSA behaviour for
$D=0$ and for $t\ll N$.
When $0<D<1$ the full coverage $\theta=1$ is attained at $t=N\log(1/D)/(1-D)$,
larger than $N$, due to the larger scattering from the occupied regions.
We may also account for {\it desorption}, simply by adding an extra negative
term (linear in $\theta$) to the rate equation (\ref{cov.eq}), effectively
decreasing the value of $D$.
The coefficient of the term may incorporate the probability of desorption
from the physisorbed or from the chemisorbed states, or both.
Allowing for desorption from the chemisorbed states would not only affect the
coverage, however, but will also imply extra terms in the rates of growth and
coalescence (eq.\ (\ref{R})).
For convenience we shall keep $D\neq 1$ in the future expressions, as it
allows us to distinguish the diffusion from the direct deposition terms.

Next we define several quantities which will be useful to examine the
morphology of the system. We shall call:
\begin{itemize}
\item $N(t)$ \ the total number of islands of occupied sites \\
\item $N(L)=N(0L0)$ \ the number of islands with $L$ sites, or number of rows
of $L$ consecutive occupied sites with at least one empty site on the left and
right \\
\item $N(L00)=N(0L00)$ \ the number of islands of $L$ sites with at
least two empty sites on the right \\
\item $N(L0L')=N(0L0L'0)$ \ the number of islands of $L$ sites separated
by an empty site from an island of $L'$ sites on the right \\
\item $N(00)$ \ the number of pairs of empty sites \\
\item $N(000)$ \ the number of trios of empty sites .
\end{itemize}
The generalization of the notation for more complicated configurations is
obvious.
The densities $n$ associated with the above numbers, are defined by
their ratio to the system size, e.g.\ $n(t)=N(t)/N$.
Since the boundary effects can be neglected, the number of occupied regions
(islands) is equal to the number of empty regions.
On average the system has left and right symmetry, so $N(L00)=N(00L)$
and $N(L0L')=N(L'0L)$.

For a $1d$ substrate, the above quantities are easily defined in terms
of the correlation functions of the local variables.
Although the system is not in equilibrium we expect it to be translationally
invariant, on average, due to the randomness of the deposition attempts.
Using this property we have, for example:
\begin{eqnarray}
n(t) & = & \sum_i \left<(1-S_i)S_{i+1}\right>/N
     \ = \ \theta - \left<S_iS_{i+1}\right> \label{nt.1} \\
n(L) & = & \sum_i
        \left<(1-S_i)S_{i+1}...S_{i+L}(1-S_{i+L+1})\right>/N \nonumber \\
     & = & \left<(1-S_0)S_1...S_L(1-S_{L+1})\right> \label{nL.1} \\
n(L00) & = & \sum_i
\left<(1-S_i)S_{i+1}...S_{i+L}(1-S_{i+L+1})(1-S_{i+L+2})\right>/N \nonumber \\
& = & \left<(1-S_0)S_1...S_L(1-S_{L+1})(1-S_{L+2})\right> \label{nL00.1} \\
n(00) & = & \sum_i \left<(1-S_i)(1-S_{i+1})\right>/N
      \ = \ 1 - 2\theta + \left<S_0S_1\right> \label{n00.1} \\
n(000) + n(1) & = & \sum_i \left<(1-S_i)(1-S_{i+2})\right>/N
      \ = \ 1 - 2\theta + \left<S_0S_2\right> \label{n000.1}
\end{eqnarray}

A number of normalization or hierarchical sum rules follow, some of which are:
\begin{eqnarray}
N(t) & = & \sum_{L=1}^{\infty} N(L) \label{nt.sum} \\
N(L) & = & \sum_{L'=0}^{\infty} N(L0L')
     \ = \ \sum_{L'=1}^{\infty} N(L0L') + N(L00) \label{nL.sum} \\
N(00) & = & \sum_{L=0}^{\infty} N(L00)
      \ = \ \sum_{L=1}^{\infty} N(L00) + N(000) \label{n00.sum} \\
N(0) \ = \ (1-\theta)N & = & \sum_{L=0}^{\infty} N(0L0)
            \ = \ N(t) + N(00) \label{n0.sum} \\
\theta N & = & \sum_{L=1}^{\infty} LN(L) \label{theta.sum} \ .
\end{eqnarray}
It is easy to show that expressions (\ref{nt.1}) and (\ref{nt.sum}) (with
$n(L)$ replaced by (\ref{nL.1})) are equivalent, using fixed or periodic
boundary condition and the property $S_i^2=S_i$.
The probability of finding an island of size $L$ and its moments,
the average island size and its mean square deviation, are therefore:
\begin{eqnarray}
P(L) & = & N(L)/N(t) \label{PL.1} \\
\left<L\right> & = & \sum_{L=1}^{\infty} LP(L) \ = \ \theta/n(t) \label{L.1} \\
\frac{\left<L^2\right>-\left<L\right>^2}{\left<L^2\right>} & = &
1 - \theta^2/\left(n(t)\sum_{L=1}^{\infty}L^2n(L)\right) \label{LL.1} \ .
\end{eqnarray}

There are three basic mechanisms by which these numbers may change with the
deposition process (where $\Delta$ denotes variation):
\begin{itemize}
\item {\it nucleation}: $\Delta N(000) = 1$ \\
\item {\it growth}: $\Delta N(L00) = - 1 \ ; \ \Delta N(L+1) = 1$ \\
\item
{\it coagulation}: $\Delta N(L0L') = - 1 \ ; \ \Delta N(L+1+L') = 1$ \ .
\end{itemize}
The rates of occurrence of each of these events per unit time are:
\begin{eqnarray}
R_n & = & \frac{N(000)}{N}  \ = \ n(000) \nonumber \\
R_g & = & \sum_{L=1}^{\infty}
          \frac{1+DL/2}{N}\left[N(L00)+N(00L)\right] \nonumber \\
    & = & 2(n(00)-n(000)) + D\sum_{L=1}^{\infty}Ln(L00) \label{R} \\
R_c & = & \sum_{L=1}^{\infty}\sum_{L'=1}^{\infty}
          \frac{1+D(L+L')/2}{N}N(L0L') \nonumber \\
    & = & n(t)-n(00)+n(000)+D\theta-D\sum_{L=1}^{\infty}Ln(L00) \nonumber \\
    & = & 1-(1-D)\theta-2n(00)+n(000)-D\sum_{L=1}^{\infty}Ln(L00) \nonumber \ ,
\end{eqnarray}
where we have used (\ref{nt.sum})-(\ref{theta.sum}).
The factor $L/2$ accounts for a particle landing and diffusing on top of
an island of size $L$ towards the right (or left).
For simplicity, we have assumed, and we will assume throughout, that in
modeling the diffusion process the random walk with traps can be replaced
by a random choice between right and left.
We shall return to this point in section \ref{s:simul}.
As expected, the sum of the rates (\ref{R}) yields the total
rate of adsorption per unit time: $\sum_i R_i=1-(1-D)\theta=Nd\theta/dt$.
The probabilities of nucleation, growth and coagulation per deposition event,
$P_n$, $P_g$ and $P_c$, are then defined by the ratios:
\begin{equation}
P_i=R_i/(Nd\theta/dt) \label{Pi} \ .
\end{equation}

It is then straightforward to write the exact equation for the rate of
variation of the number of islands, given by the difference between
the rates of nucleation and coagulation:
\begin{eqnarray}
\frac{dN(t)}{dt} & = & R_n - R_c \nonumber \\
& = & 1 - (D+1)\theta - 2n(t) + D\sum_{L=1}^{\infty} L\,n(L00) \label{nt.eq} \
,
\end{eqnarray}
where we have used (\ref{n0.sum}).
Setting $D=0$ in (\ref{nt.eq}) one obtains the RSA result
$n(t)=\theta(1-\theta)$.
Using (\ref{L.1}) and (\ref{cov.eq}), it is possible to derive from
(\ref{nt.eq}) an exact equation for the average island size $\left<L\right>$.
It is also illustrative to look at the growth of a single, isolated
island. Neglecting coagulation, the equation for the island size reads:
$dL/dt=(2+DL)/N$, which solution (with $L(0)=0$) is $L(t)=(2/D)(\exp(Dt/N)-1)$.
This, however, only gives the expected growth at early times.
When island coagulation becomes important the growth of $L$ should be faster
than exponential, and as $t\to N$ (for $D=1$) $L$ should become of order $N$.

\section{Simulation}
\label{s:simul}

We performed numerical simulations of the model described in section
\ref{s:model}. For convenience we used free boundary conditions, although the
choice of boundary conditions should be irrelevant.
The results were averaged over an ensemble comprising a great number ($N_s$) of
system samples with different random deposition histories.
For large coverages, the distribution $P(L)$ and its second moment proved
particularly sensitive to finite sampling and size effects.
This is easily understood, as the size of the larger island at late-times
(which controls the dynamics at this stage) can fluctuate by as much as $N/2$.
The smoothness of the curves for the probabilities $P_n$, $P_g$ and $P_c$ also
depends on the number of samples used.
We used $N=N_s=50000$, and verified that the systematic finite sampling and
size
errors were almost eliminated as we increased $N$ and $N_s$ up to these values.
For ease of analytical treatment (eq.\ \ref{R}) and for computational
efficiency, we modelled the precursor layer diffusion with a random
choice between right and left rather than with a random walk with traps.
Although that mechanism does not take into account the starting point of the
diffusion process, we expect this effect to be irrelevant because on average
all sites on an island are equally likely to be chosen for a deposition
attempt.

\subsection{Results}
\label{s:simres}

The measurements of $n(t)$, $1/\left<L\right>$,
$(\left<L^2\right>-\left<L\right>^2)/\left<L^2\right>$ and $P(L)$ (at six
different coverages) are displayed in Figures 1 to 4 (the pair correlations
are shown in section \ref{s:thres}).
We have divided the second moment by $\left<L^2\right>$ rather then
$\left<L\right>^2$ since the latter diverges.

It is interesting to observe the extent to which the RSA kinetics of deposition
is modified by the additional diffusion mechanism.
For comparison, we included in Figures 1 to 3 the plots (broken lines) of:
$\theta(1-\theta)$, $(1-\theta)$ and $\theta/(1+\theta)$, and in Figure 5 the
plot of $\theta^{L-1}(1-\theta)$, which correspond to the same quantities
in the RSA model.
We see from Figures 1 to 3, that the number of islands is smaller, and the
average island size and island size fluctuations are larger with diffusion
mediated deposition. This results from the increase in the rates of
island growth and coagulation relative to the rate of island nucleation.
Comparing Figures 4 and 5 for the island size distribution, we can see a
close agreement at small coverages, when all islands are still small.
At intermediate coverages, the distribution spreads out (larger islands)
in the diffusion case. At large coverages the difference is even
greater. The RSA curve becomes uniform (all island sizes are equally
probable), while the diffusion curve shows that the majority of islands
are still small in size, but there is a minority of very large islands.
At the late stages of diffusion mediated growth, the number of large islands
is small and their size can fluctuate enormously, therefore the
smoothness of the $P(L)$ curve depends strongly on the sampling number.

We have also examined the mechanisms responsible for the structure kinetics.
Figure 6 shows the evolution of the probabilities of nucleation, growth
and coagulation per deposition event, $P_n$, $P_g$ and $P_c$ (eq.\ (\ref{Pi})).
For comparison we have included the corresponding RSA curves (broken lines):
$P_{n,RSA}=(1-\theta)^2$, $P_{g,RSA}=2\theta(1-\theta)$ and
$P_{c,RSA} = \theta^2$ (cf.\ (\ref{R})-(\ref{Pi}) with $D=0$).
The most obvious difference occurs as $\theta\to 1$.
In the RSA case the probability of growth vanishes. In the diffusion case,
however, the probability of growth tends to a finite value $1-P_c$, showing
that a considerable number of gaps with two or more empty sites still exist
at large coverages.

\section{Theory}
\label{s:theor}

Due to the diffusion on top of the occupied regions, the probability of
adsorption at an empty site depends on the state of occupation of its
neighbouring sites and even of far located sites, if the site is at the edge
of a large island.
These correlations develop in the system as the coverage increases with
time and diffusion mediated deposition becomes more likely.
As a result, the density numbers defined in section \ref{s:model}
obey an infinite set of hierarchical coupled equations which cannot be solved
exactly (an example of an equation in the top of the hierarchy is given by
(\ref{nt.eq})). One must, therefore, look for approximate solutions by
truncating the hierarchy with some closure scheme.
The simplest approximation consists in incorporating only the {\it pair
correlations} in the system, which are then determined self-consistently.
We shall see that, by carefully choosing the multi-site functions decoupling
scheme, such a simple approach, despite neglecting multiple correlations,
is capable of capturing many qualitative and quantitative features of the
system's behaviour.

Let us denote the pair correlations between sites with different separations,
as:
\begin{eqnarray}
p \ = \ q^{(1)} & = & \left<S_iS_{i+1}\right> \nonumber \\
q \ = \ q^{(2)} & = & \left<S_iS_{i+2}\right> \label{pq.def} \\
q^{(n)} & = & \left<S_iS_{i+n}\right> \ , \ n\ge 1 \nonumber \ .
\end{eqnarray}
We expect the correlations to decay with distance, i.e.:
\begin{equation}
p \ge q \ge q^{(3)} \ge ... \ge \theta^2 \ . \label{decay}
\end{equation}

The quantities depending on the local variables of {\it two sites} only,
can be expressed exactly in terms of the pair correlations.
{}From (\ref{nt.1})-(\ref{n000.1}) we have:
\begin{eqnarray}
n(t)  & = & \theta - p \label{nt.2} \\
n(00) & = & 1 - 2\theta + p \label{n00.2} \\
n(000) + n(1) & = & 1 - 2\theta + q \label{n000.2} \ ,
\end{eqnarray}
and the average island size (\ref{L.1}) is given by
\begin{equation}
\left<L\right> = \frac{\theta}{\theta-p} \label{L.2} \ ,
\end{equation}

To write the {\it multi-site} functions approximately, in terms of the pair
correlations, we decouple the higher order correlators into a product chain
of pair correlators, each associated with an adjacent bond, as follows:
\begin{eqnarray}
\left<S_1S_2...S_nS_{n+m}\right> & \simeq & \frac{\left<S_1S_2\right>
\left<S_2S_3\right>...\left<S_{n-1}S_n\right>\left<S_nS_{n+m}\right>}
{\left<S_2\right>...\left<S_{n-1}\right>\left<S_n\right>} \nonumber \\
& = & \frac{p^{n-1}q^{(m)}}{\theta^{n-1}} \ \ , \ \ (n\geq 2, m\geq 1) \ .
\label{pair}
\end{eqnarray}
The normalization factors in the denominator assure that the RSA result is
recovered in the decoupling limit. Then, from
(\ref{nL.1})-(\ref{n000.1}) we have:
\begin{eqnarray}
n(L) & \simeq & \frac{p^{L-1}}{\theta^{L}}(\theta-p)^2 \ , \ (L\geq 1)
\label{nL.2} \\
n(000) & \simeq & 1 - 3\theta + 2p + q - p^2/\theta \ . \label{n000.3} \\
n(L00) & \simeq & \frac{p^{L-1}}{\theta^{L+1}}(\theta-p)
\left[\theta^2-\theta(p+q)+p^2\right] \ , \ (L\geq 1) \label{nL00.2}
\end{eqnarray}
Note that the limits of ($n(00)-n(L)$) and ($n(000)-n(L00)$) when $L\to 0$,
which are $(p-\theta^2)/p$ and $(p-\theta^2+\theta(q-p))/p$, respectively,
although non-zero have a small value.
Using these expressions and (\ref{nt.2})-(\ref{n00.2}), we can then
write the deposition rates (\ref{R}) as:
\begin{eqnarray}
R_n & = & 1 - 3\theta + 2p + q - p^2/\theta \nonumber \\
R_g & = & (D+2)(\theta-p) + 2(p^2/\theta-q) + D\theta\frac{p-q}{\theta-p}
\label{R.appx} \\
R_c & = & Dp + q - p^2/\theta - D\theta\frac{p-q}{\theta-p} \nonumber \ ,
\end{eqnarray}
and the mean square deviation of the island sizes (\ref{LL.1}) as:
\begin{equation}
\frac{\left<L^2\right>-\left<L\right>^2}{\left<L^2\right>} \ = \
\frac{p}{\theta+p} \ . \label{LL.2}
\end{equation}

For the theory to be self-consistent, we must check that the expressions
for the densities, although approximated are properly normalized.
In fact, summing (\ref{nL.2}), with $L$ from 1 to $\infty$, gives $\theta-p$
(cf.\ (\ref{nt.sum}) and (\ref{nt.2})). Also, the sum
$\sum_{L=1}^{\infty}Ln(L)$
gives $\theta$ (cf.\ (\ref{theta.sum})).
Hence, the probability distribution for the island sizes (\ref{PL.1}),
obtained by dividing (\ref{nL.2}) by $n(t)$,
\begin{equation}
P(L) = (p/\theta)^{L-1}\left(1-p/\theta\right) \ , \label{PL.2}
\end{equation}
is normalized to 1.
This is a geometric distribution, as in the RSA case, but with $\theta$
replaced by $p/\theta$.
Adding (\ref{n000.3}) and (\ref{nL00.2}), with $L$ from 1 to $\infty$, gives
$1-2\theta+p$ (cf.\ (\ref{n00.2}) and (\ref{n00.sum})).
Other more complicated density numbers turn out to be consistently normalized
too, as their sums yield the correct density number within the pair
approximation. This is the case for $n(L0L')$, which satisfies the sum
rule (\ref{nL.sum}).

The next step of our approach is to determine the pair correlations
self-consistently.
There is an infinite hierarchy of coupled equations for the pair correlations,
even within the pair approximation. The first two are the equations for $p$ and
$q$. We will now derive the exact form of these two equations.

The $p$ equation follows immediately from equation (\ref{nt.eq}) for $n(t)$,
using (\ref{nt.2}) and (\ref{cov.eq}). It is more instructive, however,
to write it down by inspection.
$pN$ is the average number of pairs $11$, of two neighbouring occupied sites,
which increases by one with island growth and by two with coalescence. Hence,
using (\ref{R}), we have:
\begin{eqnarray}
N\frac{dp}{dt} & = & 2R_c + R_g \nonumber \\
& = & 2(D+1)\theta - 2p - D\sum_{L=1}^{\infty}Ln(L00) \ . \label{p.eq.exact}
\end{eqnarray}
Within the pair approximation, using (\ref{R.appx}) and (\ref{cov.D=1}), and
setting $D=1$, we obtain:
\begin{equation}
\frac{dp}{d\theta} = 3\theta - p - \theta\frac{p-q}{\theta-p} \ .
\label{p.eq.appx}
\end{equation}

The exact equation for $q$, although more complicated can also be written
down by inspection.
$qN$ is the average number of pairs $1-1$, of two occupied sites separated by
a site in any state.
It can increase by 1 or 2 with island nucleation, growth or coalescence.
A careful consideration of all possibilities leads to the following equation:
\begin{eqnarray}
N\frac{dq}{dt}
& \!\!\!\! & = \ 2\sum_{LL'=1}^{\infty}\frac{1}{N}N(L0{\bf 0}0L') +
2\sum_{LL'=2}^{\infty}\frac{1+D(L+L')/2}{N}N(L{\bf 0}L') \label{q.eq.exact1} \\
+ & \!\!\!\! &
2\sum_{L=1,L'=2}^{\infty}\frac{1+DL'/2}{N}
\left[N(L0{\bf 0}L')+N(L'{\bf 0}0L)\right] \nonumber \\
+ & \!\!\!\! & \sum_{L=1}^{\infty} \left\{
\frac{1}{N}\left[N(L0{\bf 0}00)+N(00{\bf 0}0L)\right] +
\frac{1+D/2}{N}\left[N(L0{\bf 0}1)+N(1{\bf 0}0L)\right] \right\} \nonumber \\
+ & \!\!\!\! & \sum_{L=2}^{\infty} \left\{
    \frac{1+DL/2}{N}\left[N(L{\bf 0}00+N(00{\bf 0}L)\right] +
\frac{1+D(1+L)/2}{N}\left[N(L{\bf0}1)+N(1{\bf0}L)\right] \right\} \nonumber \ .
\end{eqnarray}
Employing the hierarchical relations (\ref{nt.sum})-(\ref{n0.sum}) and
some obvious generalizations, this equation then simplifies to
\begin{equation}
N\frac{dq}{dt} = 2 - 2(1-D)\theta - (2+D)n(1) - 2n(000) -
D\sum_{L=1}^{\infty}L\left[n(L000)+n(L010)\right] \ . \label{q.eq.exact2}
\end{equation}
Evaluating the sum of $L[n(L000)+n(L010)]$ within the pair approximation,
using (\ref{nL.2}) and (\ref{n000.3}) and putting $D=1$, yields
\begin{equation}
\frac{dq}{d\theta} = 2(\theta+p-q) -p^2/\theta +
\frac{\theta q^{(3)}-pq}{\theta-p} \ . \label{q.eq.appx}
\end{equation}
One can check that $n(L000)$ and $n(L010)$ are properly normalized within
the pair approximation (their sums give $n(000)$ and $n(1)$, respectively).
The approach also yields $\sum_{L=0}^{\infty}[n(L000)+n(L010)]=1-2\theta+q$,
which is the exact result (cf.\ (\ref{n000.2})).

It is straightforward to derive the behaviour of $p$ and $q$ for small and for
large coverage.
As $\theta\to 0$ the system is RSA like, so $p$, $q$ and $q^{(3)}$ should
behave as $\theta^2$. Equations (\ref{p.eq.appx}) and (\ref{q.eq.appx}) reduce
to $p' = 3\theta + O(\theta^2)$ and $q' = 2\theta + O(\theta^2)$, where
primes indicate derivatives with respect to $\theta$. Hence
\begin{eqnarray}
p & = & 3/2\, \theta^2 + O(\theta^3) \label{p.to0} \\
q & = & \theta^2 + O(\theta^3) \ \ , \ \ (\theta\to 0) \label{q.to0} \ .
\end{eqnarray}
The factor $3/2$ (confirmed by the simulations, Figure 7) shows that even in
this regime there is an increase in the correlations relative to the RSA case.
It results from the diffusion mediated growth of single site islands:
with RSA there 2 possibilities for growth, and with diffusion there is a
third one; therefore, there are $3/2$ as many double site islands as in RSA.
Let us now consider the limit when $\theta\to 1$.
At $\theta=1$, (\ref{p.eq.appx}) and (\ref{q.eq.appx}) yield the equations:
$(2-p'(1))(1-p'(1)) = p'(1)-q'(1)$ and
$(2-q'(1))(1-p'(1)) = q'(1)-q^{(3)'}(1)$.
The first equation and inequality (\ref{decay}) imply that
$1\leq p'(1)\leq q'(1)\leq q^{(3)'}(1)\leq ...\leq 2$. Hence, the solution is
\begin{eqnarray}
p'(1) & = & 2-(2-q'(1))^{1/2} \ = \ 2-(2-q^{(3)'}(1))^{1/3} \nonumber \\
q'(1) & = & 2-(2-q^{(3)'}(1))^{2/3} \label{pq.der} \ .
\end{eqnarray}
The actual value of the derivatives depends on the higher derivative
$q^{(3)'}$, and thus on the truncation scheme adopted.
The simulation results (Figure 7) suggest that $p'(1)=q'(1)=2$,
which implies (via (\ref{pq.der}) that $q^{(3)'}(1)=2$, and that
\begin{equation}
p \ = \ q \ = \ q^{(3)} \ = \ 2\theta - 1 + O(1-\theta)^2 \ = \
\theta^2 + O(1-\theta)^2 \ \ \ , \ \ \ (\theta\to 1) \label{pq.to1} \ .
\end{equation}
This is consistent with the truncation schemes: $q=\theta^2$ or
$q^{(3)}=\theta^2$.

\subsection{Results}
\label{s:thres}

The $p$ equation (\ref{p.eq.appx}) involves $q$;
the $q$ equation (\ref{q.eq.appx}) involves $q^{(3)}$, etc.
The nature of the approximation depends on how we close the hierarchy.
In the simplest, {\it first approximation} we neglect the correlations
beyond the nearest neighbours, i.e.\ we set $q=q_1=\theta^2$, solve numerically
equation (\ref{p.eq.appx}) for $p=p_1$, with initial condition $p(0)=0$, and
substitute $p_1$ and $q_1$ in the quantities of interest.
In the {\it second approximation} we neglect the correlations
beyond the second
 neighbours, i.e.\ we set $q^{(3)}_2=q^{(3)}=\theta^2$, and
solve the system of equations (\ref{p.eq.appx})-(\ref{q.eq.appx}) for
$p=p_2$ and $q=q_2$, with initial conditions $p(0)=q(0)=0$.

Figure 7 shows the simulation data, $p_s$ and $q_s$, and the predictions
from the first theory, $p_1$ and $q_1=\theta^2$.
The correlations from the simulation decay with distance as in (\ref{decay}),
and the differences $p_s-\theta^2$, $q_s-\theta^2$ are small but non-zero,
as they should be since they establish the difference of behaviour
relative to the RSA model. The agreement between $p_1$ and $p_s$ is quite good.
As expected, it gets slightly worse as $\theta$ approaches 1 and the
correlations between sites further apart become more relevant, but the
correct asymptotic behaviour is obtained.
Consequently, there is also good agreement in the results for the density of
islands $n(t)$ (Figure 8) and the average island size $\left<L\right>$
(Figure 9), which are (exact) functions of $p$ only.
The agreement between $q_1$ and $q_s$ (Figure 7) is, of course, less
satisfactory.
Figure 10 compares the theoretical and simulation plots for the probabilities
of nucleation, growth and coagulation, $P_n$, $P_g$ and $P_c$
(eq.\ (\ref{R.appx}) with $D=1$; cf.\ Figure 6).
There is good quantitative agreement up to $\theta=0.5$, and
there is still some qualitative agreement for larger coverages.
The theory fails, however, to give the correct asymptotic behaviour as
$\theta\to 1$: although there is a region where $P_g\simeq 1-P_c$, the
theory gives $P_g\to 0$ as $\theta\to 1$.
{}From (\ref{R.appx}) we can see that $P_g\simeq (1-q')/(1-p')-1$ close to
$\theta=1$. Hence the limit $P_g=0$ is a consequence of having $p'(1)=q'(1)$,
which follows (via (\ref{pq.der}) from the truncation $q=\theta^2$.

Nearest-neighbour correlations are sufficient to probe the presence of island
boundaries. Since the above quantities depend essentially on island counting,
their predictions are fairly accurate.
Except in the early stages of deposition, however, long-range correlations need
to be accounted for to correctly describe the spectrum of island sizes.
Hence, as a result of the pair decoupling approximation (\ref{pair}), the
theory predicts a geometric distribution for the island size probability
(eq.\ (\ref{PL.2})), the same as in the RSA case but with $\theta$ replaced by
$p/\theta$.
The plots of $P(L)$ are analogous to the RSA ones (Figure 5), though since
$p/\theta>\theta$ each curve appears to correspond to a slightly larger
coverage.
The second moment of $P(L)$ (eq.\ (\ref{LL.2})) is, of course, also incorrect:
its limit as $\theta\to 1$ is $1/2$ as in RSA, rather than 1 as in
the simulation.
{}From the plots of $P(L)$ and its second moment, we find that the theory
breaks
down for these quantities for coverages over $0.2$.

Figure 11 compares the simulation data with the predictions from the
second theory, $p_2$, $q_2$ and $q^{(3)}_2$.
As before, $p_2$ fits $p_s$ quite well, and $q_2$ fits $q_s$ even better.
Consequently, the agreement in the results for $n(t)$ (Figure 8) and
$1/\left<L\right>$ (Figure 9) is also good: there is no major difference
between the two theories, apart from the fact that now the curves lay below
the simulation plots.
A similar difference between theories is found (Figure 10) for the probability
of nucleation $P_n=n(000)$, an approximate function of both $p$ and $q$
(eq.\ (\ref{n000.3})).
A worse agreement with the simulation than in the first theory is
obtained, however, for the probabilities of growth and coagulation.
We have also tested other plausible choices for the closure scheme, as
the ones employed above are not unique, but found the results were
either largely unchanged or incorrect.

Finally, we note that $p_2$ lies over $p_s$, while $p_1$ lies below $p_s$.
$q_2$, on the other hand, lies between $q_s$ and $q_1$.
Hence, the $p$ correlations are underestimated in the first theory
and overestimated in the second theory.
This seems to indicate that the approach cannot be systematically improved by
higher order truncations in the pair correlation hierarchy.
This fact comes as no surprise given the uncontrolled nature of a
self-consistent approach. Moreover, the results are more likely to be
affected by the pair decoupling approximation (\ref{pair}) then by the
order of truncation in the pair correlation hierarchy.

\section{Conclusions}
\label{s:concl}

We have studied numerically and analytically a simple, but non-trivial
model for the deposition of monomers on a line.
The particles can diffuse on the extrinsic precursor layer until they
reach the edge of the island, where they are irreversibly deposited.
As time progresses, island nucleation becomes less frequent, while the
larger islands grow rapidly and merge with other islands.
During this process, the precursor particles migrate for larger and larger
distances with ever increasing probability, establishing correlations
between sites further and further apart.
As a result, the system develops a structure characterized by strong
correlations whose range grows to the system size at full coverage.

In the simulation, we looked at the evolving structure pattern by
measuring the island density number, the island size probability distribution
and its first and second moments, and some of the pair correlations.
We also looked at the interplay between the direct and the diffusion
mediated mechanisms of deposition, by measuring the nucleation, growth
and coalescence probabilities per unit time.
As expected, the results (Figs.\ 1-7) differ considerably from the RSA model,
especially at large coverages, due to the increasing correlations.

To explain and interpret these measurements, we proposed a simple,
self-consistent theory which incorporates correlations to a minimum extent,
i.e.\ local pair correlations.
We considered two levels of approximation, depending on the closure of
the hierarchical equations for the pair correlations.
Altogether, the lowest level of approximation, accounting for
nearest-neighbour correlations only, gave the best fit to the simulation data.
The predictions of the theory (Figs.\ 7-9) proved very accurate for the
nearest-neighbour correlator $p$, island density number $n(t)$ and average
island size $\left<L\right>$.
A fairly good prediction was also obtained for the probabilities of nucleation,
growth and coagulation (Fig.\ 10).
While the nearest-neighbour correlations are sufficient to distinguish
occupied from non-occupied regions, the full range of correlations is
required to distinguish the sizes of those regions.
An accurate determination of the island size distribution $P(L)$ and
its second moment is, therefore, beyond the scope this theory, and
would, in principle, require the use of more complicated methods.

We end with a comment on the $2d$ systems, which are of more interest
to experimentalists.
In this case, however, the same quantities are not easily expressed
in terms of the local lattice variables, and one is faced with basic
difficulties in the development of a useful formalism (most of section
\ref{s:model} would not be applicable) and in the analytical treatment.
To illustrate the problem, we note that instead of the nearest-neighbour
correlations, some non-local operator would be required to probe the domain
boundaries.
It is desirable and possible, nonetheless, to perform numerical simulations
of the $2d$ model, which would also enable the study of richer phenomena,
such as percolating clusters.

\section*{Acknowledgments}
We would like to thank Adrian Taylor for valuable discussions at the
outset of this work.

\newpage
\singlespacing

\end{document}